\documentstyle[11pt,moriond,epsfig]{article}

\def\be{\begin{equation}}
\def\ee{\end{equation}}
\def\bea{\begin{eqnarray}}
\def\eea{\end{eqnarray}}

\def\nue{\nu_{e}}
\def\num{\nu_{\mu}}

\def\nmnt{$\nu_{\mu}\leftrightarrow\nu_{\tau}$~}
\def\nenm{$\nu_{e}\leftrightarrow\nu_{\mu}$~}
\def\lsim{\lower.7ex\hbox{${\buildrel < \over \sim}$}}
\def\gsim{\lower.7ex\hbox{${\buildrel > \over \sim}$}}
\begin{document}
\vspace*{4cm}
\title{CURRENT STATUS OF THE K2K LONG-BASELINE NEUTRINO-OSCILLATION
EXPERIMENT\footnote{Talk at XXXVth  Rencontres de Moriond ``Electroweak
interactions and unified theories'', Les Arcs, Savoie, France, \\ March~11-18,~2000.
The transparencies used in this talk can be found in
{\tt http://neutrino.kek.jp/}\~{\tt oyama/public.html}}}

\author{ YUICHI OYAMA\footnote{E-mail address:~{\tt yuichi.oyama@kek.jp};
~~URL:~{\tt http://neutrino.kek.jp/}\~{\tt oyama}}\\
for K2K Collaboration\footnote{ The K2K collaboration includes about 100 physicists from KEK,
ICRR, Kobe, Niigata, Okayama, Tohoku, Tokai,
SUT, Kyoto, Boston, U.C.Irvine, Hawaii, LANL, SUNY, Washington, Warsaw, Chonnam,
Dongshin, Korea and SNU.}
\footnote{The official Web page of K2K experiment
is {\tt http://neutrino.kek.jp/}}}

\address{Institute of Particles and Nuclear Studies, \\
High Energy Accelerator Research Organization (KEK), \\
Tsukuba, Ibaraki 305-0801 Japan}

\maketitle
\vspace{5cm}
\abstracts{
The K2K (KEK to Kamioka) long-baseline neutrino-oscillation experiment
was successfully started in early 1999.
A total intensity of $7.20 \times 10^{18}$ protons
on target, which is about 7\% of the goal of the experiment, was accumulated
in 39.4~days of data-taking in 1999.
We obtained 3 neutrino events in the fiducial volume of the Super-Kamiokande
detector, whereas the expectation based on
observations in the front detectors is 12.3\hbox{${{+1.7}\atop{-1.9}}$}.
An analysis of oscillation
searches from the view points of absolute event numbers,
distortion of neutrino energy spectrum, and $\nue/\num$ ratio is in progress.}

\newpage

\section{Introduction}

The K2K experiment\cite{Nishikawa,cpvio}
is the first long-baseline neutrino-oscillation experiment
using an artificial neutrino beam. Almost a pure
$\nu_{\mu}$ beam from $\pi^{+}$ decays is generated
in the KEK 12-GeV/c Proton Synchrotron, and is detected in
Super-Kamiokande (SK) 250km away.
Neutrino oscillation can be examined from characteristics
of the neutrino events observed in SK.
The nominal sensitive region in the neutrino-oscillation parameters is
$\Delta m^{2} > 3\times 10^{-3}$eV$^{2}$, which covers the
parameter region suggested by the atmospheric neutrino
anomaly observed by several underground
experiments,\cite{Kamioka,IMB,Soudan}
and confirmed by SK.\cite{SuperK}
The sensitive regions on the neutrino oscillation parameters
for \nmnt oscillation and \nenm oscillation are shown in Figure 1. 

\begin{figure}[b!]
\centerline{
\epsfig{file=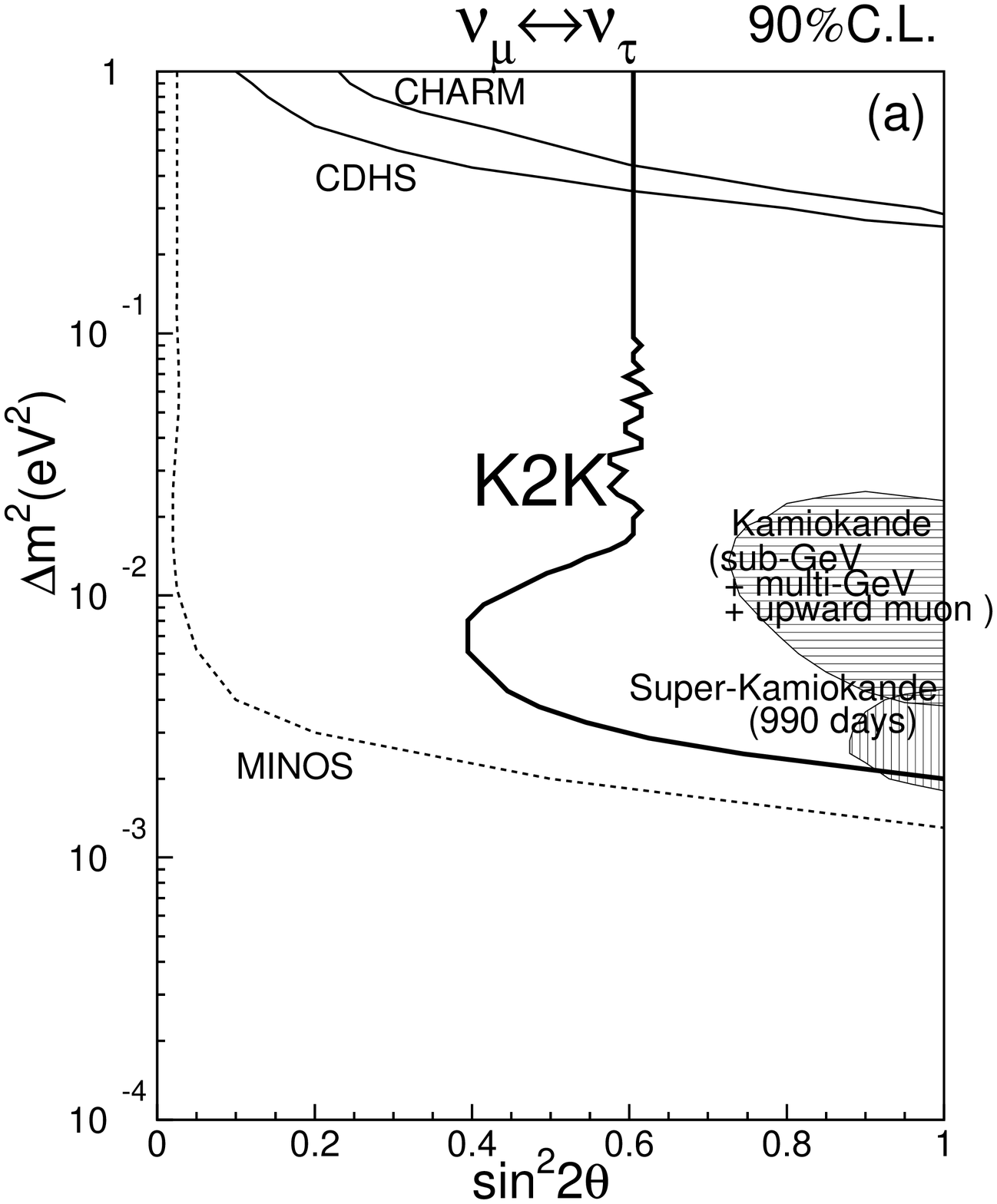,height=9.2cm}
\hskip 1.1cm
\epsfig{file=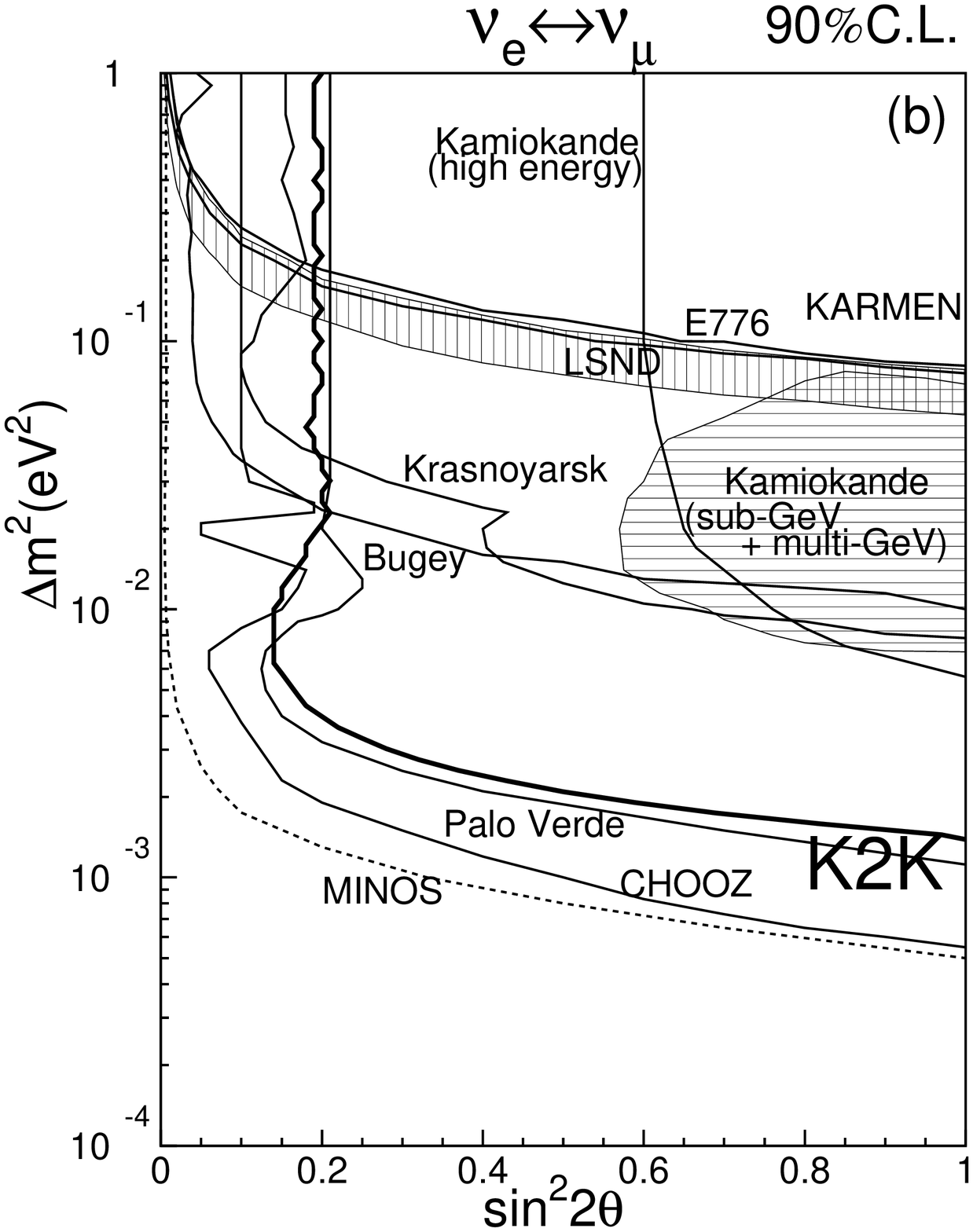,height=9.2cm}
}
\caption{
Sensitive region on the neutrino oscillation parameters,
$\Delta m^{2}$ and $\sin^{2}2\theta$ for (a)\nmnt oscillation and (b)\nenm oscillation.
The allowed region by Kamiokande,$^{3}$ LSND,$^{7}$ Super-Kamiokande,$^{8}$
and excluded (or sensitive) regions by other experiments are also plotted$^{9-19}$
in solid (dashed) lines.
The K2K calculation includes recent progress on the simulation of neutrino generation,
neutrino interaction, and detector response.~~~~~~~~~~~~~~~~~~~~~~~~~~~~~~~~~~~~~~~~~~~~
}
\label{fig:nmnt}
\end{figure}

\section{Neutrino beam and detectors}

The K2K experiment consists of (1)a proton synchrotron and neutrino beam line
including magnetic horns\cite{Horn} and various beam monitors, (2)two front detectors
(1kt water Cherenkov detector (1KT) and so-called Fine Grained detector (FGD)) 300m
downstream of the target and (3)SK as a far
detector. The detailed design and performance of those components
were already presented in Ref.2, and are not discussed in this report.
Instead, the design of the neutrino beam line and the property of the
neutrino beam are given in Table 1. The design, performance and
event rate of the detectors are summarized in Table 2. A schematic view of the
K2K front detector is also shown in Figure 2.

\begin{table}[bth]
\caption{Design of the neutrino beam line and property of the neutrino beam.}
\begin{center}
\begin{tabular}{ll}
\hline
\hline
Neutrino beam line&\\
~~$\bullet$ proton momentum  & 12 GeV/c\\
~~$\bullet$ proton intensity & 5.4 $\times 10^{12} \rm{proton/pulse}$\\
~~$\bullet$ extraction mode  & fast extraction\\
~~$\bullet$ beam duration    & $\sim 1.1~\mu$second for every 2.2 second\\
~~$\bullet$ target           & 3cm$\phi \times$ 65cm Aluminum\\
~~$\bullet$ decay tunnel     & 200 m\\
\hline
Property of the neutrino beam&\\
~~$\bullet$ mean energy      & 1.4 GeV\\
~~$\bullet$ peak energy      & 1.0 GeV\\
~~$\bullet$ $\nu_{e}/\nu_{\mu}$ & $\sim 1 \%$\\
~~$\bullet$ flux at 300m downstream    & 1.7 $\times 10^{12} \nu/{\rm cm}^{2}$ for $10^{20}$ p.o.t\\
~~$\bullet$ flux at 250km downstream   & 1.3 $\times 10^{6} \nu/{\rm cm}^{2}$ for $10^{20}$ p.o.t\\
\hline
\hline
\end{tabular}
\end{center}
\end{table}

\def\lsim{\lower.7ex\hbox{${\buildrel {+1.7}\over{-1.9}}$}}

\section{Summary of the data in 1999}

The K2K experiment was successfully started in early 1999.
The first neutrino beam was generated on January 27.
After machine studies and beam tuning, the first physics
run was started on March~3.
In 1999, about 100~days of physics data-taking was scheduled.
However, the successful physics data-taking was only for 39.4~days, due
to some problems with the neutrino beam line.
On June~19, the first neutrino event was observed in SK.
In the 1999 run, the integrated proton intensity was $7.20\times 10^{18}$ p.o.t.
(protons on target), which is about 7\% of the goal of the
experiment, $10^{20}$ p.o.t.\cite{Nishikawa} 

The neutrino beam direction has been confirmed to agree with the direction of
the SK detector within 0.3~mrad based on the data of the beam monitors.\cite{cpvio}
Because the absolute flux and energy spectrum of the neutrino beam are expected to be
almost the same within 3~mrad, the adjustment of the neutrino beam direction is sufficient.

\begin{figure}[b!]
\centerline{\psfig{file=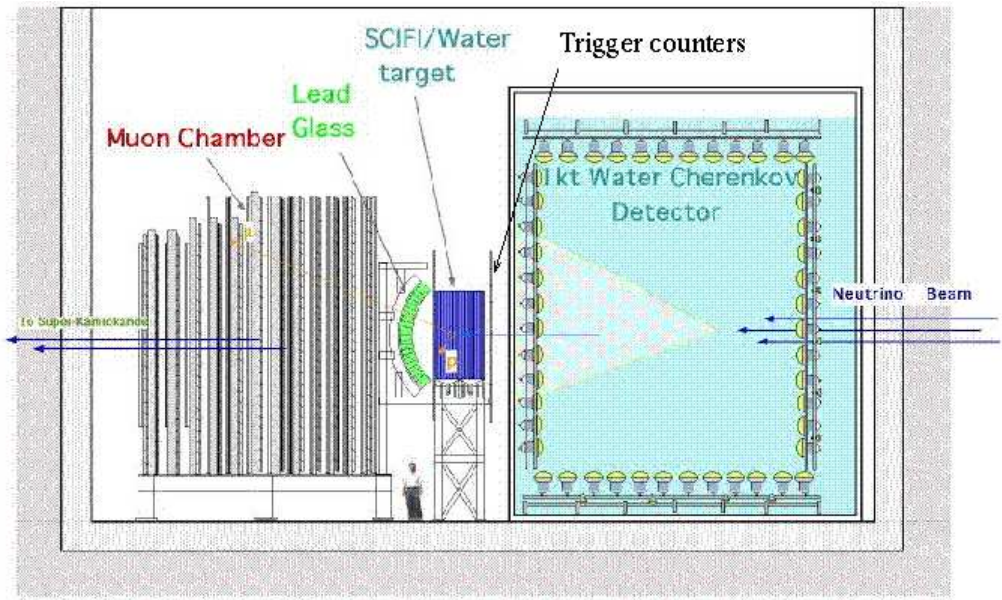,height=8.0cm}}
\caption{Front detector in the K2K experiment.
         The arrangement of the trigger counters is slightly changed
         from Ref.2}
\label{fig:neard}
\end{figure}

The selection of neutrino events in SK employs
the time difference between the neutrino beam and each event.
Considering the neutrino beam duration (1.1$\mu$sec) and accuracy of
the absolute time determination ($<0.3\mu$sec), events within a 1.5$\mu$sec
time window covering the neutrino beam period are selected.
A total of 12 events have been found, which includes events in the inner
counter, in the outer counter and interactions in the surrounding rock.
Among them, 3 events were detected in the fiducial volume of the inner counter.
Because the expected atmospheric neutrino background in the fiducial
volume within the neutrino beam period is calculated to be $2\times 10^{-4}$ events,
the 3 events in the fiducial volume are a clear signal of neutrinos
from KEK.

\begin{table}[p]
\caption{Design, purpose, performance, event rate of detectors in the K2K experiment.
Some of the numbers given in Ref.2 have been replaced with the results of measurements
~~~~~~~~~~~~~~~~~~~~~~~~~~~~~~~~~~~~~~~~~~~~~~~~~~~~~~~~~~~~~~~~~~~~~~~~~~~~~~~~~~~}
\vspace{0.4cm}
\begin{center}
\begin{tabular}{ll}
\hline
\hline
\multicolumn{2}{l}{{\bf Front detectors 300m downstream}} \\
\multicolumn{2}{l}{{\bf (A)1kt water Cherenkov Detector}} \\
~~$\bullet$ design     & 1/50 (total volume) miniature of the Super-Kamiokande detector. 496 tons of\\
                       & water are viewed by 860 20-inch $\phi$ PMTs. The fiducial mass is 50.3~ton.\\
~~$\bullet$ purpose    & direct comparison with neutrino events in Super-Kamiokande\\
~~$\bullet$ performance& $e/\mu$ identification capability:~~$>$99\%\\
                       & $\Delta E_{e} / E_{e}$:~~3\%/$\sqrt{E_{e}{\rm (GeV)}}$\\
~~$\bullet$ event rate & 0.01 events/spill in fiducial volume; 2 $\times 10^{5}$ events for $10^{20}$ p.o.t.\\

\multicolumn{2}{l}{{\bf (B)Fine Grained Detector}}\\
~~$\bullet$ design     & consists of a scintillating fiber tracker, trigger counters,\\
                       & lead-glass counters and muon chamber\\
~~$\bullet$ purpose    & precise measurement of the neutrino-flux profile and energy distribution\\

\multicolumn{2}{l}{~(B-1)scintillating fiber tracker (SFT)}\\
~~~$\bullet$ design   & 20 layers \char'134 sandwich'' of scintillating fiber (0.7mm$\phi$) sheets and water in aluminum\\
                      & containers. The sensitive area is 2.4m $\times$ 2.4m; the fiducial mass is 5.94 tons\\
~~~$\bullet$ function & track reconstruction of charged particles and identification of the kinematics\\
                      & of neutrino interactions. Water is used as the target material.\\ 
~~~$\bullet$ performance & detection efficiency of each layer:~~$>$99\%\\
                         & position resolution:~~$\sim 280 \mu$m\\
~~~$\bullet$ event rate  & 0.001 events/spill, 2 $\times 10^{4}$ events for $10^{20}$ p.o.t.\\

\multicolumn{2}{l}{~(B-2)trigger counters (TRG)}\\
~~~$\bullet$ design   & 80 large plastic scintillators (466cm (L)$\times$10.7cm(W)$\times$4.2cm(T)) \\
                      & covering upstream and downstream of the scintillating fiber tracker\\
~~~$\bullet$ function & rejection of cosmic ray muons and neutrino events from 1kt water\\
                      & Cherenkov detector, and measurement of the absolute event time \\
~~~$\bullet$ performance &  timing resolution:~~$\sim$2nsec\\
                         &  position resolution:~~$\sim$5cm\\
                         &  detection threshold@center of the scintillators:~~$\sim$1.7MeV\\
                         &  detection efficiency for a penetrating charged particle:~~$>$ 99\%\\

\multicolumn{2}{l}{~(B-3)lead glass counters (LG)}\\
~~~$\bullet$ design   & 600 lead glass counters with acceptance of 11.3cm $\times$12.2 cm each \\
~~~$\bullet$ function  & identification of electrons from the energy deposit in the counter\\
~~~$\bullet$ performance &  $\Delta E_{e}/E_{e}$:~~$\sim 10\%/\sqrt{E_{e}{\rm (GeV)}}$\\

\multicolumn{2}{l}{~(B-4)muon chamber (MUC)}\\
~~~$\bullet$ design   & 12 layers \char'134 sandwich'' of $\sim$900 drift chambers and iron filters(10cm-20cm thickness)\\
~~~$\bullet$ function  & measurement of the muon energy from the range\\
~~~$\bullet$ performance &  position resolution:~~2.2mm\\
                         &  detection efficiency of each layer:~~$\sim$ 99\%;\\
                         &  $\Delta E_{\mu}/E_{\mu}:~~8\sim 10\% $\\
\hline
\multicolumn{2}{l}{{\bf Far detector (Super-Kamiokande) 250km downstream}}\\
~~$\bullet$ design     & 50kt huge water Cherenkov detector at about 1000m underground \\
                       & 22.5kt of the fiducial volume is viewed by 11164 20-inch $\phi$ PMTs \\
~~$\bullet$ performance& $e/\mu$ identification capability:~~$>$ 99 \%;\\
                       & $\Delta E_{e} / E_{e}$:~~3\%/$\sqrt{E_{e}{\rm (GeV)}}$\\
                       & $\Delta E_{\mu} / E_{\mu}$:~~3\%\\
                       & accuracy of absolute event time adjustment with neutrino beam:~~$<$0.3$\mu$sec\\
~~$\bullet$ event rate & $\sim$0.3 events/day, $\sim$172 events for $10^{20}$ p.o.t.\\
\hline
\hline
\end{tabular}
\end{center}
\end{table}

\section{Present status of data analysis}

Strategies concerning oscillation searches at K2K are summarized as follows. 
The \nmnt oscillation can be examined by a disappearance of neutrino events
in SK because the energy of the neutrino beam is smaller
than the $\tau$ production threshold. To recognize a reduction
of neutrino events efficiently, expected event number
without oscillation must be accurately estimated
using the observed event numbers in the front detectors.

In addition, the neutrino energy spectrum in SK should be
distorted in the case of oscillation because the oscillation probability
depends on the energy of the neutrinos.
Therefore, the expected neutrino energy spectrum for no oscillation
must be calculated from an extrapolation of the spectrum in the front
detectors.

An examination of the \nenm oscillation is an appearance search.
A possible excess of electron neutrino events in SK is
direct evidence of the \nenm oscillation, because the beam from KEK is almost
pure muon neutrinos in the case of no oscillation, and because
the particle identification capability in SK
has already been proved to be excellent.\cite{SuperK,Kasuga}
In order to attain a better oscillation sensitivity,
the fraction of electron neutrinos in the original neutrino beam,
which is estimated to be about 1\% from a Monte-Carlo simulation, must be
experimentally measured as precisely as possible.
Therefore, the $\nue/\num$ ratio observed in the front detectors is a key point of
the \nenm oscillation analysis.

The following three subsections discuss the present status of the front detector
analysis along these three strategies, i.e. 
(1)absolute event number, (2)shape of the neutrino energy spectrum, and
(3)$\nue/\num$ ratio.
At present, those studies are impossible in SK because the statistics
is not sufficient.

\subsection{absolute event numbers}

The expected event numbers in SK ($N^{SK}_{exp}$) are calculated based on the observed
event numbers in the front detectors, and extrapolation of the front detector
to SK. $N^{SK}_{exp}$ is written as

$$ N^{SK}_{exp} = {{N^{FD}_{obs} \times N^{SK}_{cal}}\over{N^{FD}_{cal}}},$$

\noindent
where $N^{FD}_{obs}$ is the observed event numbers in the front detectors;
$N^{SK}_{cal}$ and $N^{FD}_{cal}$ are the calculated event numbers in
SK and the front detectors, respectively, using the same simulation program. 
We employed neutrino interactions in 50.3~tons fiducial volume of 1KT
for $N^{FD}_{obs}$ and $N^{FD}_{cal}$.
In addition, neutrino interactions in the scintillating
fiber tracker\cite{Scifi} and the muon chamber were also used
to examine the consistency between the observations in the front detectors.
The results are summarized in Table 3.
$N^{FD}_{obs}/N^{FD}_{cal}$, which shows an agreement
of observed event number with the simulation, is found to be
$0.84 \sim 0.85$ for three independent observations, and is
consistent with each other.

The expected event number in SK is calculated to be
12.3\hbox{${{+1.7}\atop{-1.9}}$} in 22.5ktons of the fiducial volume, and $\sim$31
events in the total volume. Although the observed event numbers, 3 in fiducial volume
and 12 in the total volume, are considerably smaller than the expectations,
nothing can be concluded about the neutrino oscillation at this stage
because of poor statistics.

\begin{table}[b!]
\caption{Summary of the absolute event number calculation. The event numbers in the total volume
in SK means events in inner counter, outer counter, and interactions in the surrounding rock.
The error for all volume is still under study.}
\begin{center}
\begin{tabular}{lcrcll}
\hline
\hline
detector  & Fiducial~~~~   & Event~~~  & $N^{FD}_{obs}/N^{FD}_{cal}$ & \multicolumn{2}{l}{Event numbers in SK}\\
        & mass (ton) & number &   & (fid. vol.) & (total vol.)\\
\hline
\hline
1kt water Cherenkov    & 50.3 & 17672   & 0.84 $\pm$ 0.08 & 12.3\hbox{${{+1.7}\atop{-1.9}}$} & $\sim$31\\
Scintillating fiber tracker          & 5.94 & 662     & 0.85\hbox{${{+0.08}\atop{-0.09}}$}~~~ &  &  \\
muon chamber   & 445  & 56062   & 0.85 $\pm$ 0.11&                                  &     \\
\hline
\multicolumn{4}{l}{Super-Kamiokande data} & ~3 & ~~12\\
\hline
\hline
\end{tabular}
\end{center}
\end{table}

\begin{figure}[t!]
\centerline{
\epsfig{file=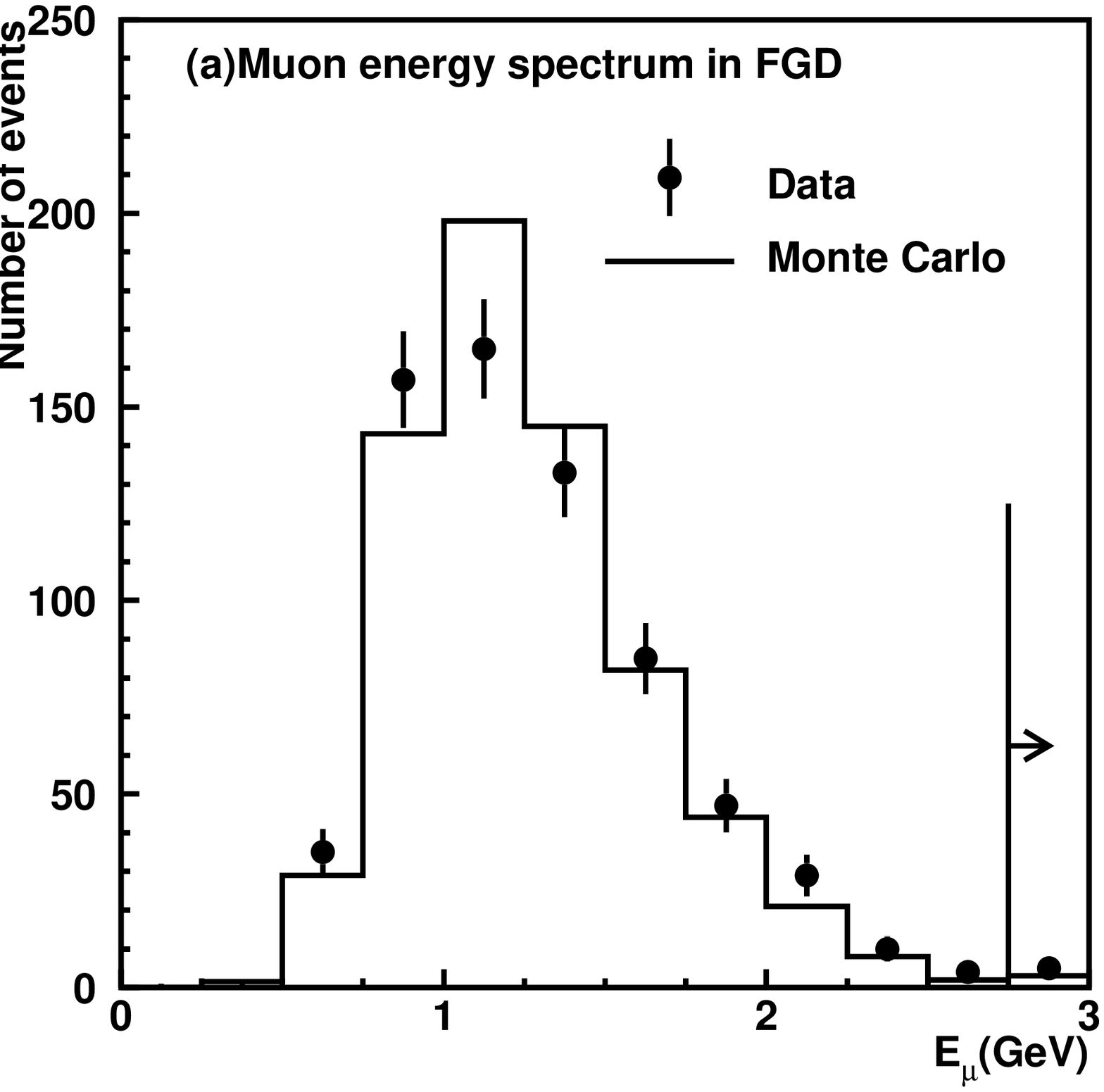,height=7.0cm}
\epsfig{file=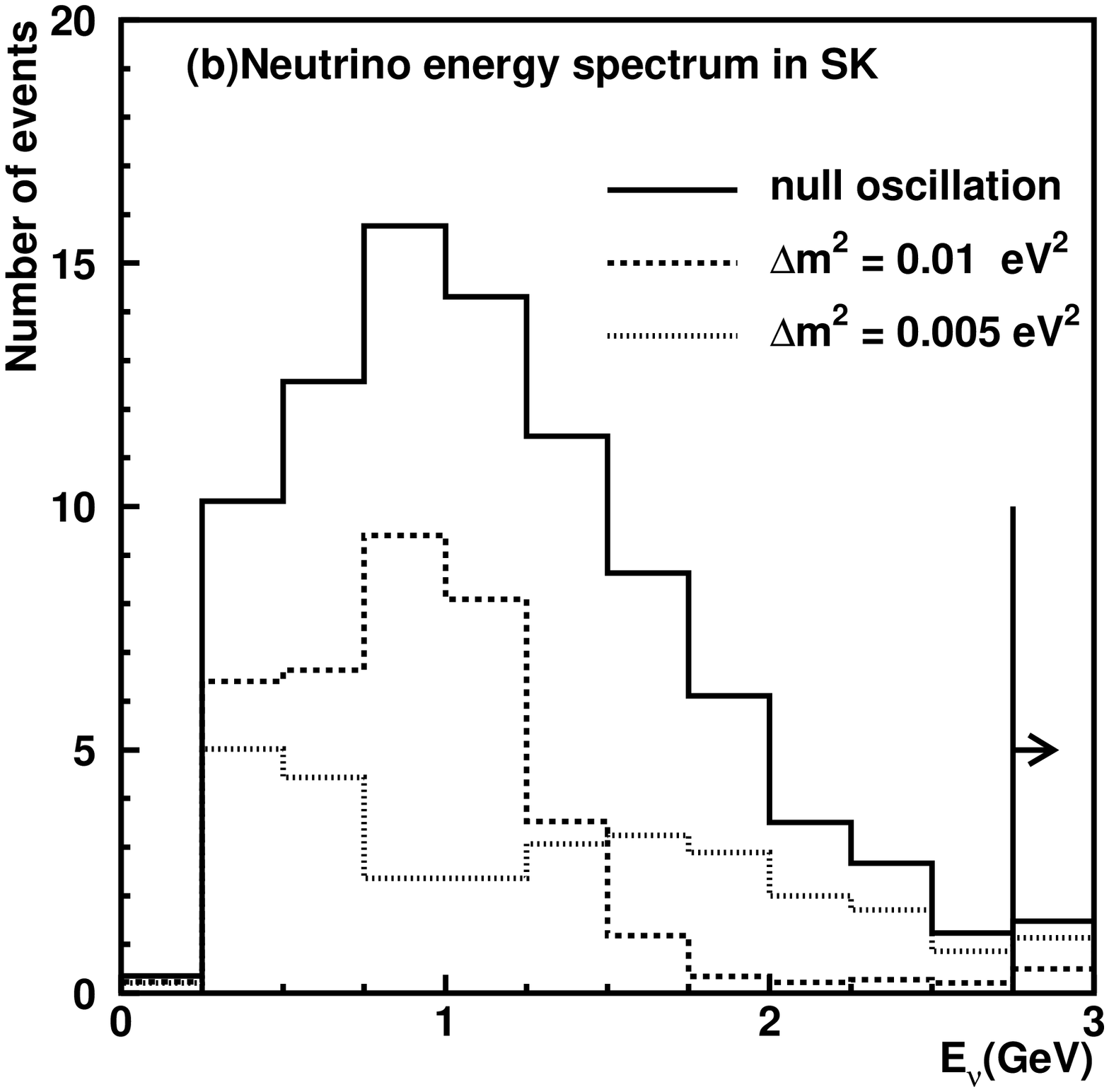,height=7.0cm}
}
\caption{(a)Moun energy spectrum obtained
         from the Fine-Grained detector. The expectation from
         the neutrino flux simulation is also shown.
	 (b)Expected neutrino energy spectrum in Super-Kamiokande.
         The spectra for the given oscillation parameters are also shown
         in dashed line ($\Delta m^{2} = 0.01$eV$^{2}$ and $\sin^{2}2\theta = 1$) and dotted line
         ($\Delta m^{2} = 0.005$eV$^{2}$ and $\sin^{2}2\theta = 1$). The shape of the energy spectrum
         in the case of oscillation is obviously distinguishable with
         the spectrum for no oscillation.
         ~~~~~~~~~~~~~~~~~~~~~~~~~~~~~~~~~~~~~~~~~~~~~~~~~~~~~~~~~~~~~~~~~~~~~~
         ~~~~~~~~~~~~~~~~~~~~~~~~~~~~~~~~~~~~~~~~~~~~~~~~}
\label{fig:spectrum}
\end{figure}

\subsection{neutrino energy spectrum}

To determine the neutrino energy spectrum,
quasi-elastic interactions of muon neutrinos, $\nu_{\mu}N \rightarrow \mu N'$,
in the scintillating fiber tracker are employed.
This is because most of the neutrino energy is transfered to the muons in
quasi-elastic interactions, and the muon energy can be measured from the range in
the muon chamber.
The neutrino energy can be directly calculated from the muon energy
with a small correction related to the scattering angle of the muon.
It should also be noted that quasi-elastic scatterings
are detected as single ring events in SK, and can be easily analyzed.

The muon energy distribution for quasi-elastic
interactions in the scintillating
fiber tracker is shown in Fig.3-(a) along with expectations
from a Monte-Carlo simulation.
For a comparison, the expected neutrino energy spectrum in SK
is shown in Fig.3-(b) together with the 
spectrum for two sets of oscillation parameters, $\Delta m^{2}=0.01$eV$^{2}$
and $\Delta m^{2}=0.005$eV$^{2}$.
The shape of the muon energy spectrum in Fig.3-(a) agrees with the spectrum from a
Monte-Carlo simulation, and is somehow correlated with the
expected neutrino energy distribution shown in Fig.3-(b).
However, a calculation of neutrino
energy spectrum from the muon energy distribution, and its extrapolation to SK
is still being studied. 

\subsection{$\nue/\num$ ratio}

The $\nue/\num$ ratio in the front detectors are
independently measured using 1KT and FGD.

As reported in Ref.21, the e/$\mu$ identification analysis
in water Cherenkov detectors employs a likelihood function which
quantitatively evaluates the agreement of the Cherenkov ring patterns with
electrons and with muons. The likelihood distribution for the neutrino beam
obtained in the 1KT is shown in Fig.4-(a)
together with an expectation from
a Monte-Carlo simulation. The likelihood distribution for atmospheric
neutrino interactions obtained by SK is shown in
Fig.4-(b) for a comparison. In Fig.4-(b), two peaks
corresponding to electron neutrinos and muon neutrinos
can be clearly distinguished.
On the other hand, in Fig.4-(a), most of the Cherenkov
ring patterns are judged to be muons, and the agreement with the distribution
for a Monte-Carlo simulation is excellent.
Although the $\nue/\num$ ratio experimentally obtained in 1KT is found to be
as small as the Monte-Carlo expectation, the numerical results are
still being analyzed.

\begin{figure}[t!]
\centerline{
\epsfig{file=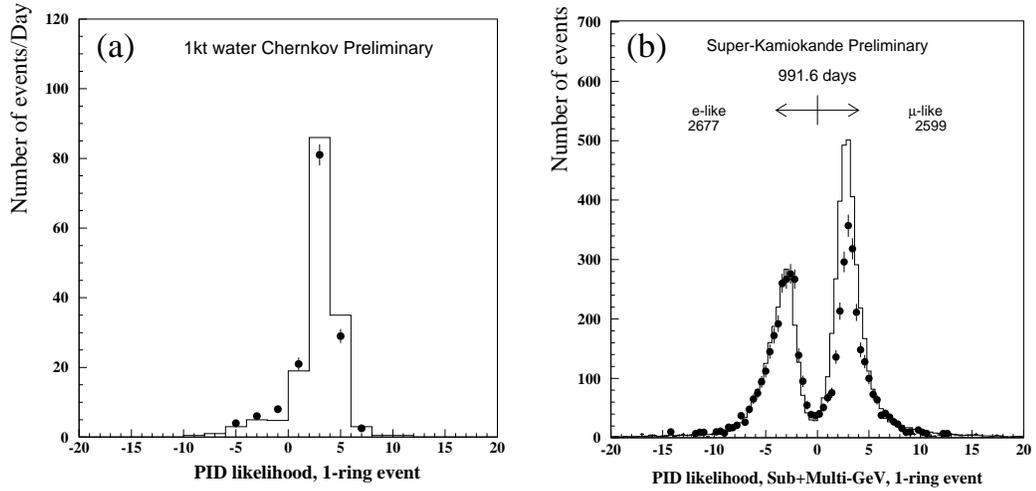,height=7.0cm}
}
\caption{Particle identification likelihood distribution
for (a)neutrino beam measured by the 1KT,
and for (b)atmospheric neutrino obtained by SK.
If the likelihood is positive (negative), the event is judged as a muon
(an electron). The expectation from Monte Carlo simulations are also
shown.
~~~~~~~~~~~~~~~~~~~~~~~~~~~~~~~~~~~~~~~~~~~~~~~~~~~~~~~~~~~~~~~~~~~~~~~~~~~~~~~~~~~~~~~~~~~~~~} 
\label{fig:likelihood}
\end{figure}

\begin{figure}[b!]
\centerline{
\epsfig{file=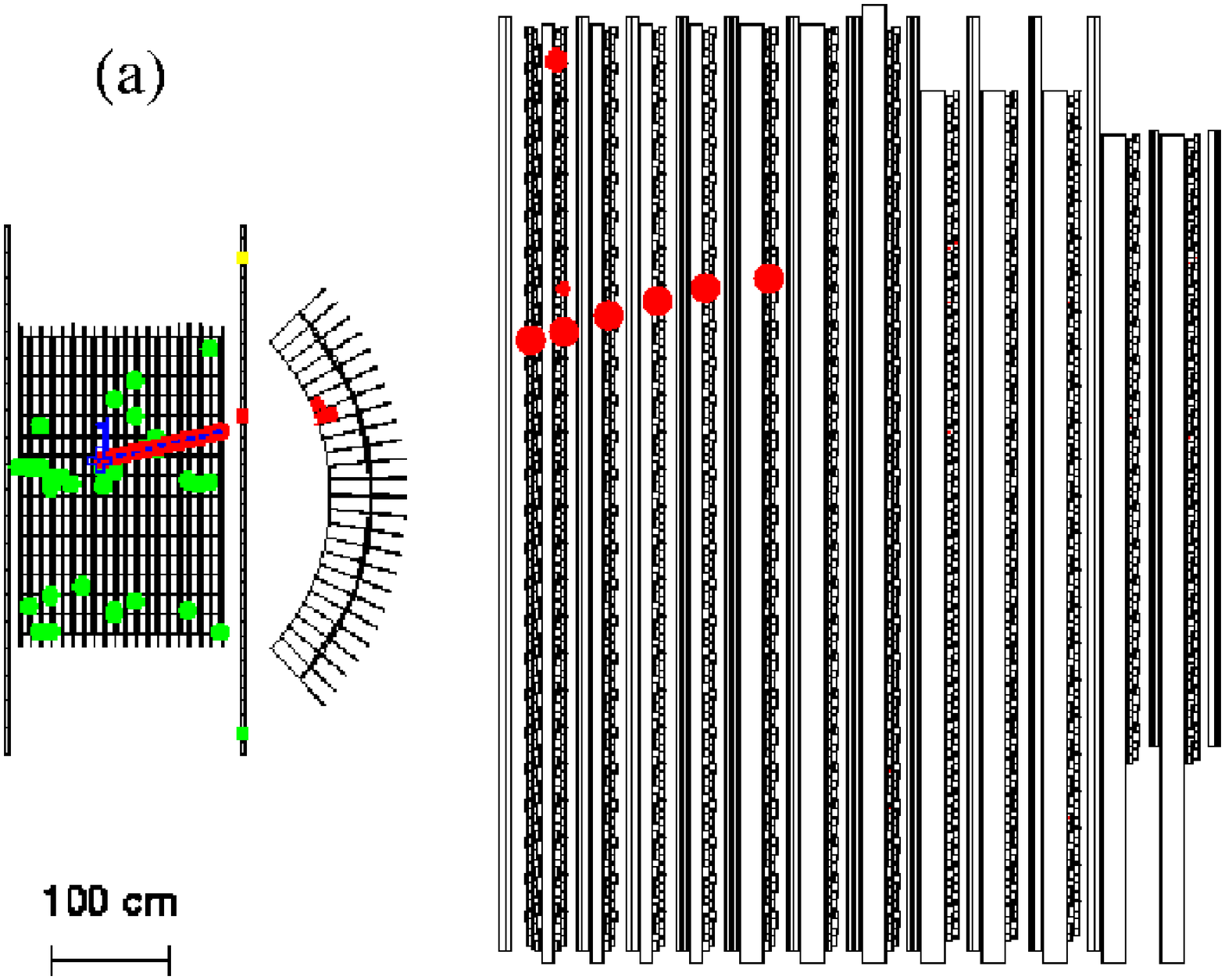,height=6.2cm}
\hskip 1cm
\epsfig{file=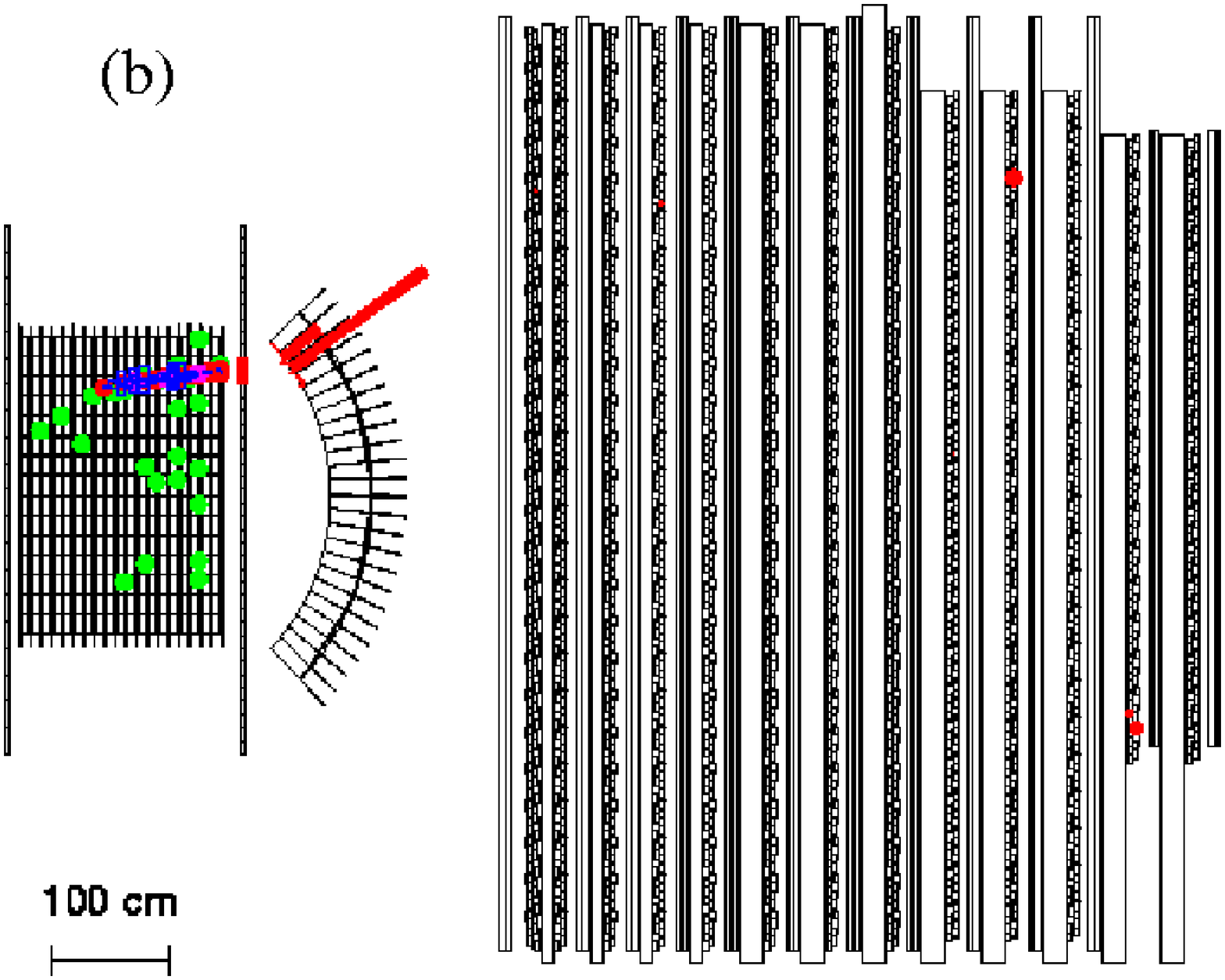,height=6.2cm}
}
\vskip 1cm
\centerline{
\psfig{file=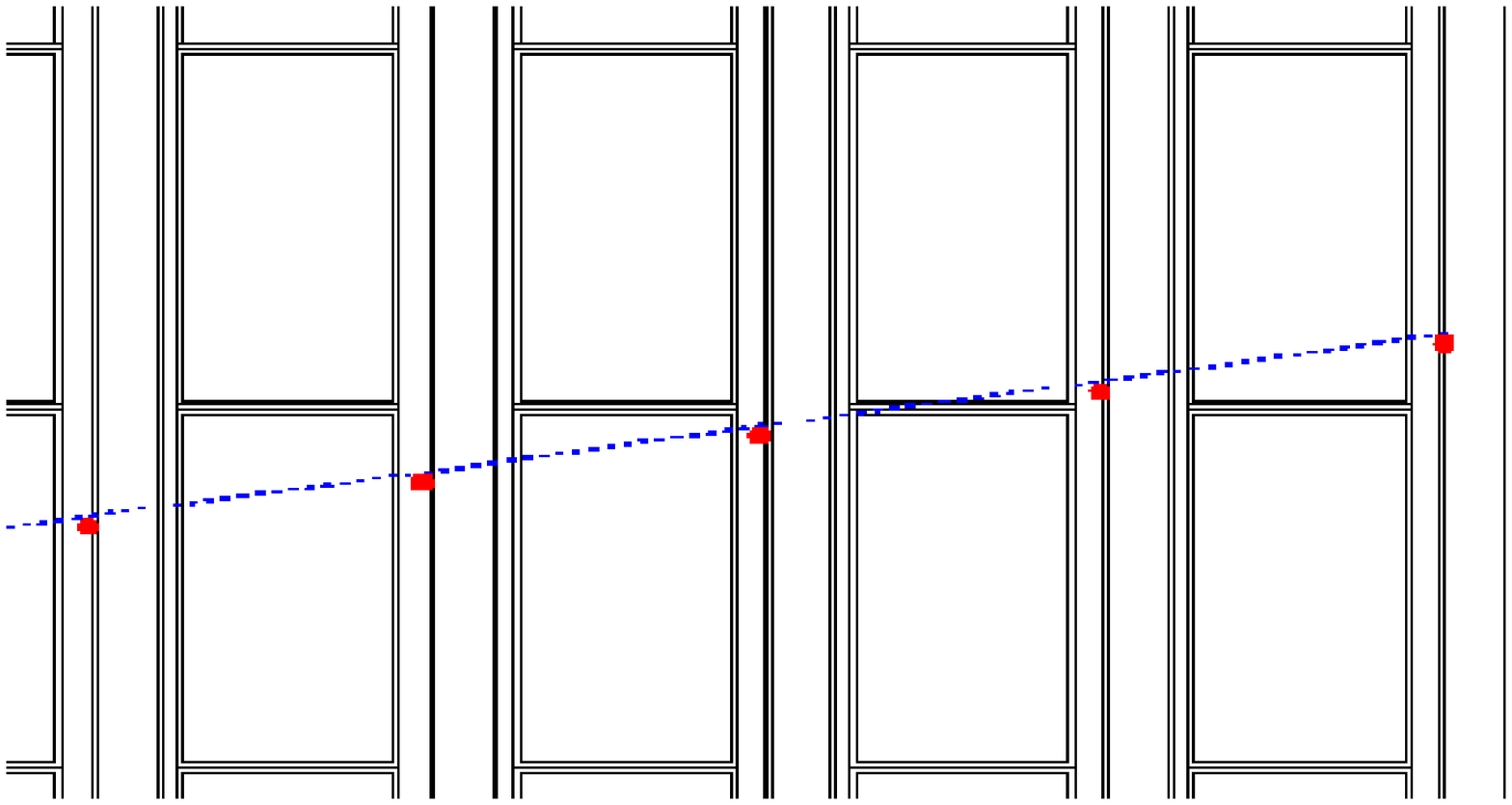,height=3.3cm}
\hskip 2.5cm
\psfig{file=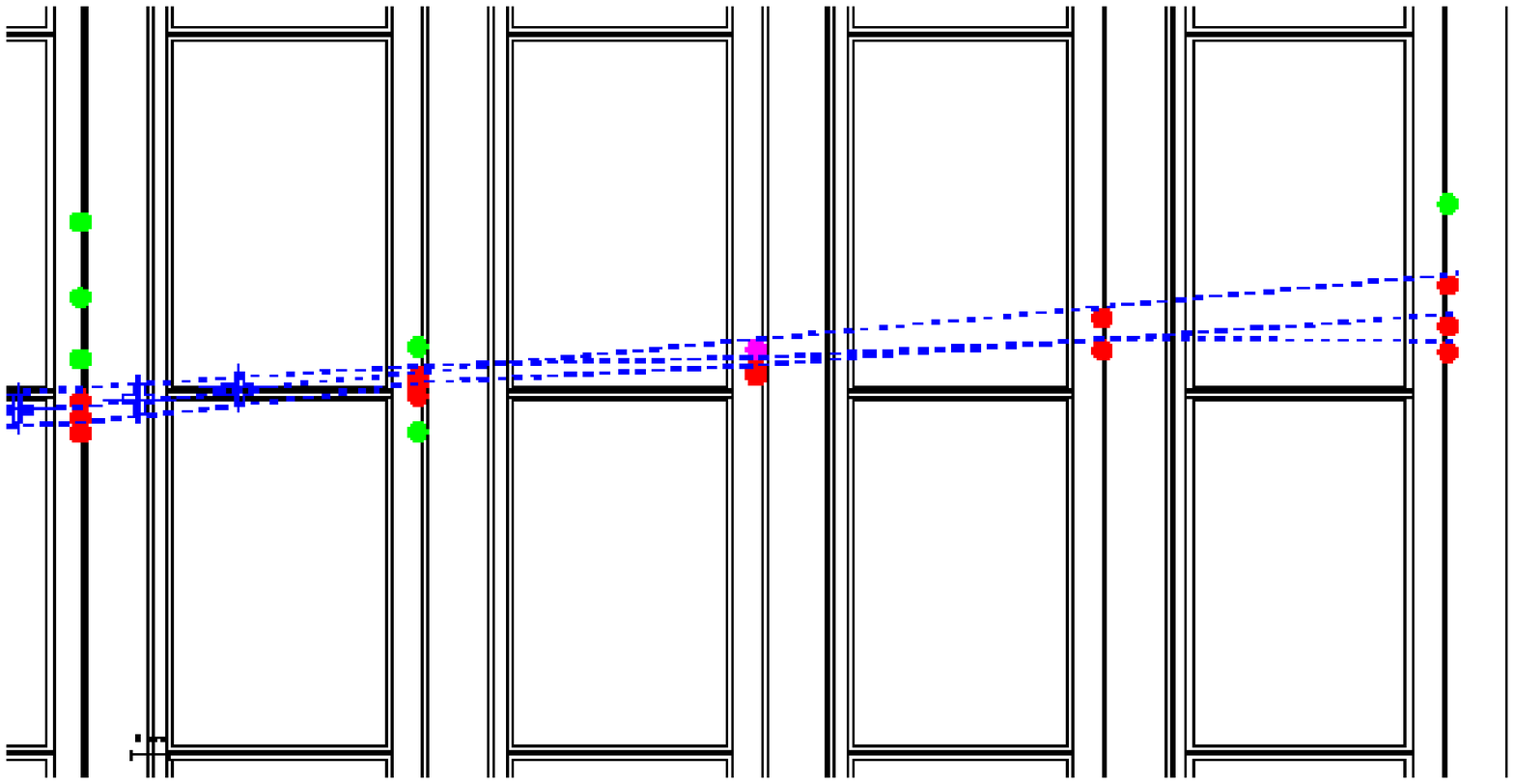,height=3.3cm}
}
\caption{(a)Typical muon neutrino event, and (b)an electron neutrino event in FGD.
         Expanded views of the scintillating fiber
         tracker are also shown at the bottom.
~~~~~~~~~~~~~~~~~~~~~~~~~~~~~~~~~~~~~~~~~~~~~~~~~~~~~~~~~~~~~~~~~~~~~~~~~~~~~~~~~~~~~~~~~~~~~~~~~~~~}
\label{fig:event}
\end{figure}

The identification of an electron neutrino event in the FGD
uses the response of each detector component for electrons
and muons. A typical muon neutrino event and an electron neutrino event
are shown in Fig.5-(a) and Fig.5-(b), respectively.
In Fig.5-(a), a muon is generated in the water target of the scintillating
fiber tracker, and produces a clear and long single muon track.
The muon penetrates the trigger counters and the lead-glass counters, and
is stopped in the middle of the muon chamber.
In Fig.5-(b), on the other hand, the electron produces an electromagnetic
shower in the scintillating fiber tracker; three charged particles
are generated as shown in the expanded view of Fig.5-(b).
These charged particles lose all of their energies
in the lead glass counters by producing a further electromagnetic
shower. No particles escape from the lead-glass counters.
Considering these characteristics, electron neutrino events are
identified by (1)a large track multiplicity in the scintillating fiber tracker,
(2)a large energy deposit in the lead glass counters, and (3)an absence
of tracks in the muon chamber. Specially, the selection using the
lead glass counters is efficient to separate muons and electrons
because the energy deposit from a single muon in the lead-glasses
is found to be 450$\pm$150 MeV from cosmic-ray muon
data, and energy resolution for 1~GeV electron is
about 10\% from an electron beam test. At present, the selection criteria
for the electromagnetic shower events is being tuned. 

Although a small $\nue/\num$ ratio is indicated by both 1KT
and FGD, the numerical results on the $\nue/\num$ ratio are still being studied.

\section{Summary and future prospect}

During 39.4 days of successful data-taking, a total intensity of
$7.20 \times 10^{18}$ protons on target were accumulated in 1999.
In Super-Kamiokande, we observed 3 neutrino interactions in the fiducial
volume, where the expectation is 12.3\hbox{${{+1.7}\atop{-1.9}}$}.
An oscillation analysis focusing on the absolute event number,
the distortion of the neutrino energy spectrum,
and the $\nue/\num$ ratio is in progress.

From January 2000 to March 2001, about 160 days of physics data taking are scheduled.
If the data can be accumulated with 100\% efficiency, we will obtain a total intensity of
$46 \times 10^{18}$ p.o.t., and about 70 events will be accumulated by end
of March 2001.

\end{document}